\documentclass[10pt,twoside,a4paper,twocolumn]{article}

\usepackage[utf8]{inputenc}
\usepackage{graphicx}
\usepackage{dcolumn}
\usepackage{bm}
\usepackage{authblk}
\usepackage{amssymb}
\usepackage{amsmath}
\usepackage{flushend}
\usepackage{placeins}  
\usepackage{hyperref}
\usepackage{abstract}
\usepackage{arydshln}  
\usepackage{color} 

\makeatletter
\renewcommand*{\@fnsymbol}[1]{\ensuremath{\ifcase#1\or *\or\or \dagger\or \ddagger\or
\mathsection\or \mathparagraph\or \|\or **\or \dagger\dagger
\or \ddagger\ddagger \else\@ctrerr\fi}}
\makeatother

\newcommand{\E}{\textsc{e}}

\date{\vspace{-5ex}}

\begin{document}

\title{Convexity constraints on linear background models\\ for electron energy-loss spectra}%

\author[,a]{Wouter Van den Broek\thanks{E-mail address: wouter.vandenbroek@thermofisher.com.}\footnote{\copyright \ This manuscript version is made available under the CC-BY-NC-ND 4.0 license \url{https://creativecommons.org/licenses/by-nc-nd/4.0/}}}%
\author[b,c]{Daen Jannis}%
\author[b,c]{Jo Verbeeck}%

\affil[a]{Thermo Fischer Scientific, Achtseweg Noord 5, 5651 GG Eindhoven, Netherlands.}
\affil[b]{Electron Microscopy for Materials Research (EMAT), University of Antwerp, Groenenborgerlaan 171, 2020 Antwerp, Belgium.}
\affil[c]{Nanolab center of excellence, University of Antwerp,  Groenenborgerlaan 171, 2020 Antwerp, Belgium.}

\twocolumn[

\maketitle

\paragraph{Abstract}
In this paper convexity constraints are derived for a background model of electron energy loss spectra (EELS) that is linear in the fitting parameters. The model outperforms a power-law both on experimental and simulated backgrounds, especially for wide energy ranges, and thus improves elemental quantification results. 
Owing to the model's linearity, the constraints can be imposed through fitting by quadratic programming. This has important advantages over conventional nonlinear power-law fitting such as high speed and a guaranteed unique solution without need for initial parameters. As such, the need for user input is significantly reduced, which is essential for unsupervised treatment of large data sets.
This is demonstrated on a demanding spectrum image of a semiconductor device sample with a high number of elements over a wide energy range.


\ \\

]\saythanks

\section{Introduction}

The rising importance of spectral imaging places strong demands on the underlying data processing tools.  To satisfy the need to quickly process large numbers of spectra, and because the amount of data far surpasses the ability of a human operator to carefully vet or fine-tune individual spectrum results, fast and robust methods are needed.

In electron energy loss spectroscopy (EELS), elemental abundances are determined by counting the number of electrons which have undergone an element-specific inelastic interaction. For example, oxygen abundance is determined by measuring the number of electrons which have inelastically scattered with the K-shell electrons. One of the most critical steps in obtaining this number of electrons is to remove the background signal. A standard method is fitting a power-law, $AE^{-r}$, to an energy window before the onset energy of the core-loss edge of interest. This method has several disadvantages: there is a user bias by defining the fitting window; a large enough pre-edge region needs to be available, which is often not the case when core-loss edges are nearby in energy; the power-law is non-linear, hence the numerical optimization can get stuck in local minima, and proper starting parameters are necessary; finally, the statistical noise is amplified due to the extrapolation underneath the core-loss edge \cite{egerton1996_44}.

Multiple methodologies have been developed to circumvent one or multiple of these issues. For instance, the extrapolation problem is solved by interpolation of the background if the edge extends to high enough energy losses \cite{Egerton_2002,Held_2020}; the non-linear aspect of the background fitting and extrapolation is removed by using a linear background function, which is a sum of fixed power-laws\cite{cueva2012}.

Another methodology is the model-based approach \cite{leapman_1988,Manoubi_1990, verbeeck_2004}, where the background and the core-loss edges are fitted to a single model which describes the physical parameters of an experimental EEL spectrum. Hence the model both includes the background and the edges, removing the extrapolation problem and the user bias since no energy windows for pre-edge fitting needs to be determined. Moreover, this approach also helps in the separation of overlapping core-loss edges\cite{leapman_1988} and takes the multiple scattering into account if the zero-loss peak is available.

In this work, we concentrate on the background signal of core-loss spectra. This consists of the contributions of many energy-loss processes that happen at onset energies below the energy offset of the spectra, such as plasmons or other core-loss edges; these have high energy tails which approximate power-laws~\cite{egerton1996_36, egerton1996_44}. 

In practice the exponent of these processes is not well known and their shape will be further modified by multiple scattering. However, despite these confounding factors, one can expect monotonically decreasing and convex behavior due to the power-law shape of the individual contributions; this holds even in the presence of multiple scattering, since the convolution with the low-loss spectrum that models it preserves monotonic decrease and convexity.

If one of the contributions to the measurements happens at energies below the spectrum onset, but at high-enough energy to not yet behave in a decreasing, convex fashion---think, for example, of the fine structure of a core-loss edge whose onset lies just a few tens of~eV below the spectrum offset---it cannot be considered a background signal within the framework of this paper, but would need to be modeled separately.

This paper investigates a background model that is linear in the fitting parameters, implements it in the model-based approach,
and formulates constraints that ensure the background's convexity, monotonic decrease and non-negativity.

The associated constrained least squares problem is completely linear and can  hence be solved fast with quadratic programming (QP) optimization. Owing to the convexity of the error metric, its single solution is obtained without the need for a starting guess.

The conventional background model in EELS is a power-law \cite{egerton1996}, the exponent of which needs fitting as well. It is this single parameter that turns the optimization problem non-linear.  In this paper, we consider the linear background model,
\begin{eqnarray}
  bg(E) = \sum_{i=1}^n a_i E^{-r_i}, \label{eq:bg}
\end{eqnarray}
with $E$ the energy-loss, $a_i$s the fitting parameters and $r_i$s fixed exponents set by the user. A similar model has been treated in~\cite{cueva2012}. Fitting the proposed model as-is, can lead to non-physical backgrounds that exhibit so-called `shoulders', are locally increasing, or even negative.  

In this paper we show that these artifacts are overcome by imposing the aforementioned constraints on the solution with QP, and that the constraints can be formulated as a relatively small number of expressions that are linear in the the fitting parameters. When applied to experimental spectra, it was observed that the model in (\ref{eq:bg}) describes the actual background better, \emph{i.e.} with lower chi-squared adjusted for degrees of freedom, and that it enables fits over larger energy regions than the conventional power-law model.

The paper is laid out as follows.  In Sec. \ref{sec:met} the methods are described, with particular attention to quadratic programming in Sec. \ref{sec:quapro}; the formulation of the constraints in Secs. \ref{sec:concon} and \ref{sec:furcon}; and the validation of the linear background model in Sec. \ref{sec:vallin}. Experimental results are shown in Sec. \ref{sec:expres}, and Secs. \ref{sec:dis} and \ref{sec:con} contain the discussion of the results and the conclusions.

\section{Methods}
\label{sec:met}

\subsection{Quadratic programming}
\label{sec:quapro}

In quadratic programming (QP), a problem of the form 
\begin{eqnarray} 
  \arg \min_x & \frac{1}{2} x^T P x + q^T x, \\
	\text{subjected to} & Gx \leq h, \\
	& Ax = b,
\end{eqnarray} 
is solved for $x$.  If the problem is convex, i.e. $P$ is positive semidefinite, it is sometimes not much more difficult to solve than a linear program  \cite{nocedal1999}.

A least squares problem can be converted into a QP, as
\begin{eqnarray}
		\arg \min_x \frac{1}{2} ||Rx - s||^2_W,
\end{eqnarray}
is equivalent to,
\begin{eqnarray}
  P = R^T W R, \ q = -R^T W s.
\end{eqnarray} 
Here, $W$ is a 
symmetric
weight matrix (usually diagonal) and $s$ is the vector of observations.  In our case, the vector of unknowns contains the parameters $a$ of the background model in (\ref{eq:bg}), and the amplitudes of the ionization edges.

The QP is solved with the Quadprog Python package \cite{quadprog}, which is based on the paper in~\cite{goldfarb1983}.

\subsection{Convexity constraints}
\label{sec:concon}

A function is convex if the line connecting any two points that lie above its graph, lies above the graph between the two points. The twice differentiable functions that are the subject of this paper, are convex if and only if their second derivatives are non-negative on the energy interval under consideration. The convexity constraint is needed, because without it, the monotonic decrease and non-negativity constraints still allow for shoulders in the background fit. 

Demanding convexity amounts to imposing a positive second order derivative with respect to the energy $E$,
\begin{eqnarray}
  \sum_i a_i s_i E^{\Delta r_i} \geq 0 \quad \forall E \in [E_b, E_e], \label{eq:2ndder}
\end{eqnarray}
with $E_b$ and $E_e$ the first and last value in the energy-axis,%
hence $E_b < E_e$, 
and $s_i = r_i (r_i + 1)$ and $\Delta r_i = r_1 - r_i$
were defined to make the notation lighter. Without loss of generality, the exponents $r_i$ are sorted in ascending order. 

In this formulation, as many constraints as energy bins are given.  To be of more practical use, this number must be reduced.

\subsubsection{Necessary conditions}
\label{sec:neccon}

It is fairly straightforward to formulate necessary conditions by evaluating (\ref{eq:2ndder}) in only a few energy values $E_j$,
\begin{eqnarray}
  \sum_i a_i s_i E_j^{\Delta r_i} \geq 0. \label{eq:necess}
\end{eqnarray}
The set of energies $E_j$ are chosen by the user.  In practice only a few energies seem to suffice: three to five are often enough.

In order to accommodate the much larger slope at lower energies, we chose to space the $E_j$s such that each is the harmonic mean of its two nearest neighbors at lower and at higher energy, and to set the lowest $E_j$ equal to $E_b$ and the highest to $E_e$.
It has to be noted however, that these conditions do not guarantee convexity.

\subsubsection{Sufficient conditions}
\label{sec:sufcon}

In this Section, (\ref{eq:2ndder}) is condensed in a lower number of constraints, while still guaranteeing convexity.

Consider the parameters $(a_1, \ldots, a_n)^T$, and move the negative terms to the right:
\begin{eqnarray}
  \sum_{i \in \mathcal{P}} a_i s_i E^{\Delta r_i} \geq -\sum_{j \in \mathcal{N}} a_j s_j E^{\Delta r_j}. \label{eq:split}
\end{eqnarray}
with $\mathcal{P} = \{k | a_k \geq 0 \}$, the set of indices of the non-negative parameters, and $\mathcal{N} = \{k | a_k < 0 \}$ that of the negative parameters.

\begin{figure}
\includegraphics[width=0.99\linewidth]{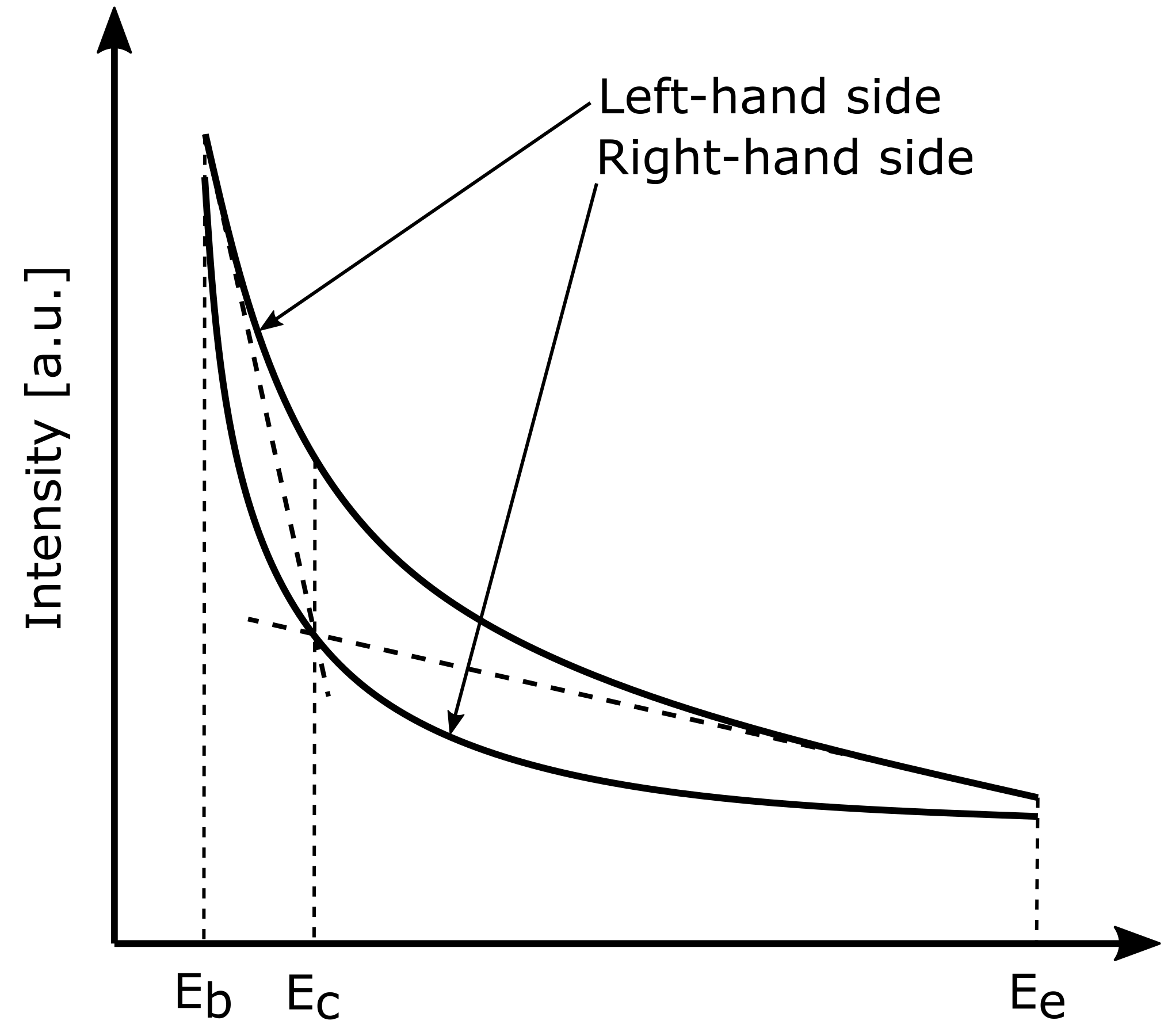}
\caption{Illustration of the role of the tangents in the convexity constraints in (\ref{eq:split}). \label{fig:tangent}}
\end{figure}

Both sides of this expression are positive, monotonically decreasing, and convex, thus ensuring that (\ref{eq:split}) holds for all energies if (\emph{i}) the right-hand side is below the left-hand side in $E_b$ and $E_e$, and (\emph{ii}) if in $E_c$ the right-hand side is lower than the left-hand side's tangent through $E_e$; where $E_c$ is defined as the crossover of the left-hand side's tangents through $E_b$ and $E_e$.
A proof of this lemma is provided in Appendix \ref{sec:conpro}.
Expressed mathematically, we get,
\begin{eqnarray}
  \sum_{i \in \mathcal{P}} a_i s_i E_b^{\Delta r_i} \geq -\sum_{j \in \mathcal{N}} a_j s_j E_b^{\Delta r_j},& \label{eq:suff1}\\
  \sum_{i \in \mathcal{P}} a_i s_i E_e^{\Delta r_i} \geq -\sum_{j \in \mathcal{N}} a_j s_j E_e^{\Delta r_j},& \label{eq:suff2}\\
  \begin{split}
  \label{eq:suff3}
  \sum_{i \in \mathcal{P}} & a_i s_i E_e^{\Delta r_i} \left( 1 + (E_c/E_e - 1) \Delta r_i \right)\\
  & \geq -\sum_{j \in \mathcal{N}} a_j s_j E_c^{\Delta r_j}.
  \end{split}
\end{eqnarray}
This is illustrated in Figure \ref{fig:tangent}.

So far this exercise was performed for one particular realization of positive and negative parameters $a_i$. However, all possible combinations need to be analyzed, and those constraints that are equivalent must be identified and eliminated. This yields the system of linear constraints,
\begin{eqnarray}
  \sum_i a_i s_i E_b^{\Delta r_i} \geq 0, \quad & \label{eq:suff_a}\\
  \sum_i a_i s_i E_e^{\Delta r_i} \geq 0, \quad & \label{eq:suff_b}\\
  \begin{split}
  &\sum_{i \in \mathcal{C}_k} a_i s_i E_e^{\Delta r_i} \left( 1 + (E_c/E_e - 1) \Delta r_i \right) \\
  & + a_1 s_1 + \sum_{i \notin \mathcal{C}_k} a_i s_i E_c^{\Delta r_i} \geq 0, \ \forall \mathcal{C}_k. \label{eq:suff_c}
  \end{split}
\end{eqnarray}
The set $\mathcal{C}_k$ is a combination of $k$ elements from the set $\{2, 3, \ldots, n \}$, and the sums are over all $k$-combinations for all $k$ ranging from $1$ to $n-2$.  The number of constraints then equals $2^{n-1}$.

Fortunately, $E_c$ can be well-approximated independently of the fitting parameters $a_j$. For a single term $a_i s_i E^{\Delta r_i}$ in (\ref{eq:split}), the crossover of the tangents is given as 
\begin{eqnarray}
  E_{c,i} = \frac{\Delta r_i - 1}{\Delta r_i} \frac{E_b^{\Delta r_i} - E_e^{\Delta r_i}}{E_b^{\Delta r_i - 1} - E_e^{\Delta r_i - 1}}.
\end{eqnarray}
Lowest values are reached for largest $|\Delta r_i|$, and can be used as a safe lower bound.  Furthermore, for typical values of $\Delta r_i$, $E_b$ and $E_e$ ($-4$, $200$~eV and $1000$~eV, respectively), $E_c$ varies relatively little as a function of $\Delta r_i$.  This can be understood by considering the limit for large $E_e$:
\begin{eqnarray}
  \frac{\Delta r_i - 1}{\Delta r_i} E_b.
\end{eqnarray}

In Appendix \ref{sec:loose} looser sufficient conditions are shown; although they guarantee convexity, they exclude
a larger number of valid solutions.

\subsubsection{Non-negative coefficients}
\label{sec:nonneg}

Another way of ensuring convexity is through the constraints
\begin{eqnarray}
  a_j \geq 0, \ \forall j.
\end{eqnarray}
These conditions are also sufficient, although they exclude notably more acceptable solutions than the constraints in Section \ref{sec:sufcon}.

\subsection{Further constraints}
\label{sec:furcon}

The constraints in Sections \ref{sec:neccon} and \ref{sec:sufcon} only impose convexity.  By adding the constraint 
\begin{eqnarray}
  \sum_i a_i r_i E_e^{\Delta r_i} \leq 0,
\end{eqnarray}
the solution becomes monotonically decreasing, and by adding
\begin{eqnarray}
  \sum_i a_i E_e^{\Delta r_i} \geq 0,
\end{eqnarray}
it is guaranteed to be positive or zero.

Note that the non-negativity constraint from Section \ref{sec:nonneg} does not need these extra conditions to guarantee monotonic decrease and non-negativity of the background.

Since physically the edge amplitudes cannot become negative, these too have been constrained to non-negative values with QP.

\subsection{Validation of the linear model}
\label{sec:vallin}

To get a sense of scale for the value of the exponent in the power-law model, it is fitted to a large number of experimentally obtained backgrounds. From various spectra with varying noise content, energy ranges of variable length were extracted that contained no edges and only background.  From these ranges, as many non-overlapping sections as possible of lengths ranging from 150~eV to 1500~eV were selected for further evaluation. 

In \cite{egerton1996bg}, it is discussed that the various contributions to the background to inner-shell edges have an approximate power-law behavior with the exponent usually close to $-3$. Indeed, fitting a power-law to the experimental background regions shows that the exponent has a mean value of $2.7 \pm 0.9$; see Table~\ref{tab:r}. This immediately suggests reasonable values and range for the exponents $r_i$ in the linear model.  

\begin{table}
\center 
\begin{tabular}{ | r | r | c | c | }
   \hline
   $\Delta E$ & \# & $\langle r \rangle$ & $\sigma_r$ \\
   \hline
   150 & 1650 & 3.62 & 0.77\\
   200 & 1650 & 3.48 & 0.76\\
   300 & 940 & 3.29 & 0.80\\
   500 & 518 & 3.04 & 0.95\\
   700 & 390 & 3.11 & 0.90\\
   1000 & 122 & 2.71 & 0.82\\
   1500 & 32 & 1.18 & 0.03\\
   \hline
\end{tabular}
\caption{Average exponent $\langle r \rangle$ of power-laws fitted to a series of experimental backgrounds over various energy windows ($\Delta E$~[eV]). The number of spectra per energy range is indicated as \#, and the standard deviation on $r$ as $\sigma_r$.} 
\label{tab:r} 
\end{table}

\subsubsection{Simulated backgrounds}

\begin{figure}
\centering
\includegraphics[width=0.99\linewidth]{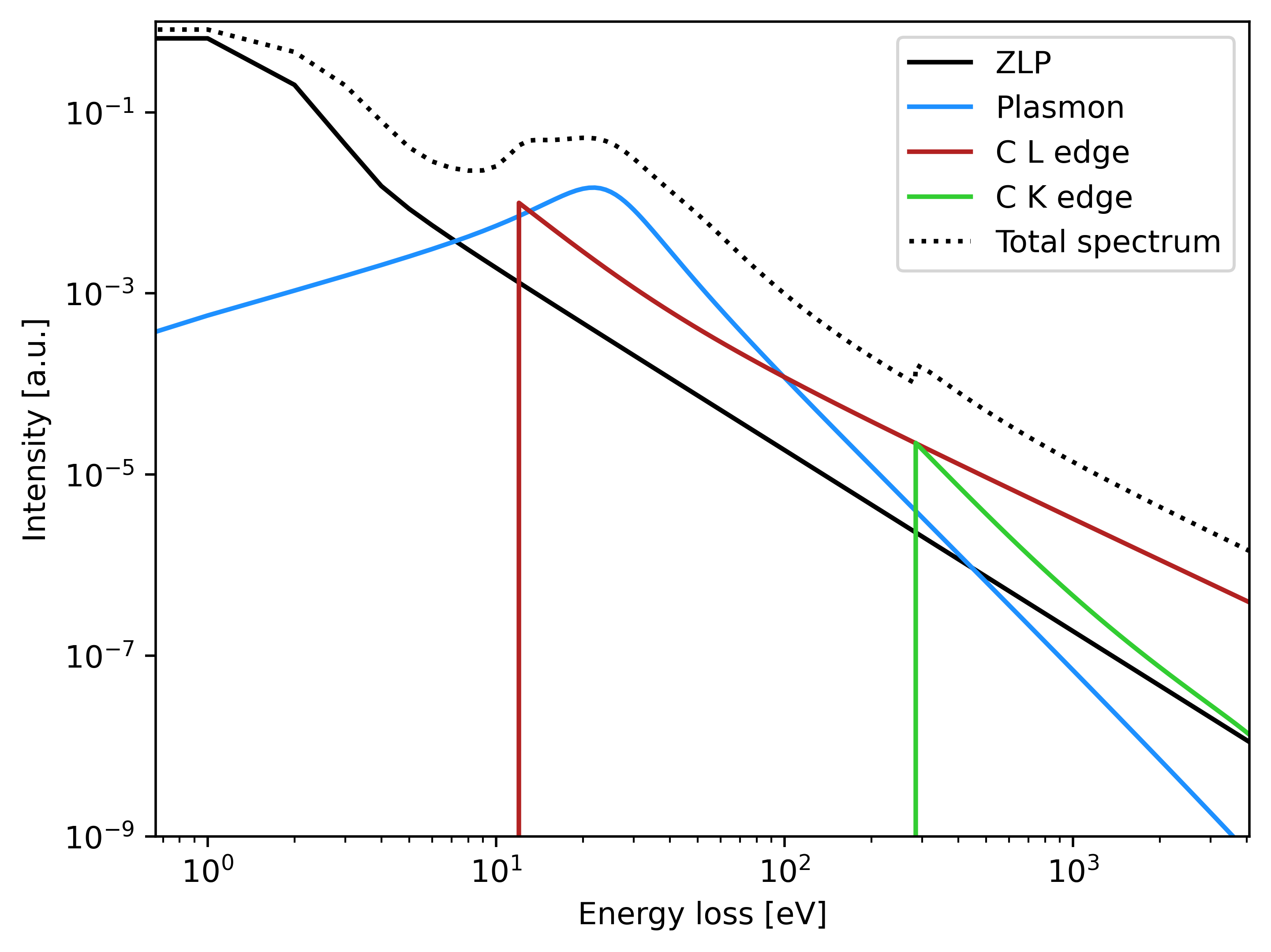}
\caption{The individual components that, after addition and self-convolution, yield the simulated background spectrum. \label{fig:simbg}}
\end{figure}

To test the different models, simulated (and hence noise free) backgrounds were created from the [400,~1900]~eV energy interval of a simulated pure carbon sample. The simulated spectrum consists of a zero-loss peak that is approximated as a Voigt function \cite{temme2022} with a full width at half maximum of 2~eV, a plasmon peak corresponding to a sample thickness of 50~nm (inelastic cross section from \cite{egerton1996}), and hydrogenic K- and L-edges with onsets at 283~eV and 11~eV \cite{Haris2017}, respectively; see  Figure \ref{fig:simbg}. The L-edge is used to mimic the single electron excitation\cite{Menon_2002}.

The dispersion was chosen at 1~eV, and the spectrum was convolved with itself to account for multiple scattering. Fine structure was not considered since the energy loss far from the edge onsets, which is the background area of interest here, are considered to be described very well by the free atom approximation.

From this 1500~eV-wide energy range, as many non-overlapping sections as possible of lengths given in the first column of Table \ref{tab:simbg} were selected for fitting with a five-term linear background, with non-negative coefficients, sufficient and necessary constraints. The exponents are given as,
\begin{eqnarray}
    r_1, \ldots, r_5 = 1, \ 2, \ 3, \ 4, \ 5. \label{eq:ri5}
\end{eqnarray}

The fit quality is assessed through the average relative error defined as, 
\begin{eqnarray}
    \langle \sigma_{\text{rel}} \rangle = \frac{1}{N} \sum_i^N \frac{|f_i - g_i|}{g_i}. \label{eq:relerr}
\end{eqnarray}
With $N$ is the number of energy bins, and $f$ and $g$ are the fitted and simulated backgrounds, respectively.

\begin{table}
\center 
\begin{tabular}{ | c | r | r | c | c | c | c |}
   \hline
   $\Delta E$ & \# & pwr & n-n & suff & nec \\
   \hline
150 & 10 & 0.05 & 1.7\E-3 & 1.9\E-4 & 1.5\E-4 \\
200 & 7 & 0.09 & 2.2\E-3 & 4.8\E-4 & 2.9\E-4 \\
300 & 5 & 0.19 & 4.8\E-3 & 1.4\E-3 & 1.1\E-3 \\
500 & 3 & 0.56 & 1.5\E-2 & 1.9\E-3 & 1.9\E-3 \\
700 & 2 & 1.22 & 2.7\E-2 & 2.0\E-3 & 2.0\E-3 \\
1000 & 1 & 3.55 & 5.1\E-2 & 5.5\E-3 & 5.5\E-3 \\
1500 & 1 & 5.66 & 1.4\E-1 & 3.2\E-2 & 2.2\E-2 \\
 \hdashline
1500 & 1 & 14.92 & 5.8\E-1 & 1.7\E-1 & 1.1\E-1 \\
   \hline
\end{tabular}
\caption{$100 \times \langle \sigma_{\text{rel}} \rangle$ [\%] of a power-law (pwr), and a five-term linear model with non-negative coefficients (n-n), sufficient (suff), and necessary (nec) constraints, fitted to the simulated backgrounds. The last row lists the maximum relative error over the interval. The number of spectra per energy range is indicated as \#.} 
\label{tab:simbg} 
\end{table}

\begin{figure}
\centering
\includegraphics[width=0.99\linewidth]{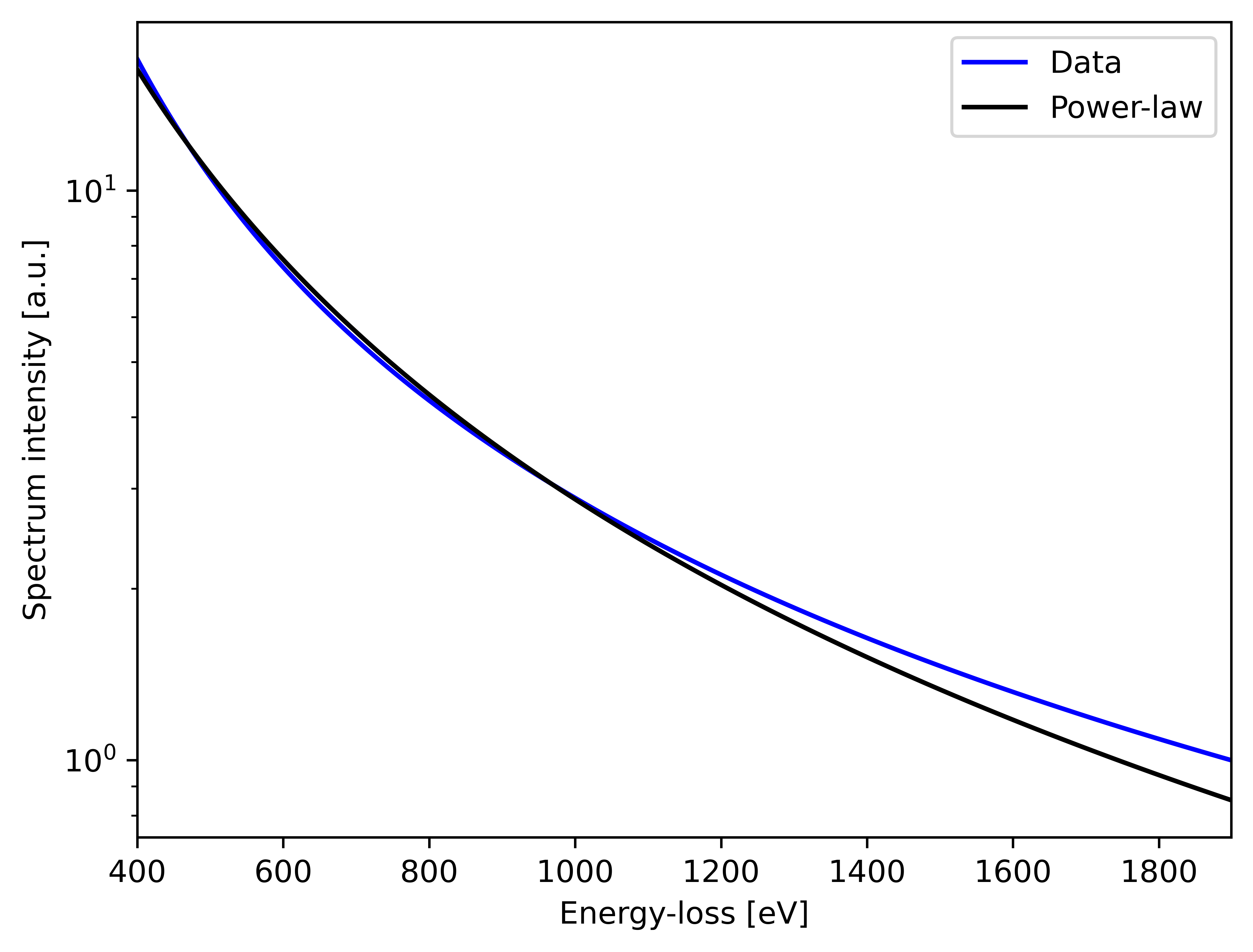}
\caption{Fit of the power-law to a 1500~eV wide simulated background. The other fits are not displayed because they are indistinguishable from the data at the plot's display resolution; see Figure \ref{fig:relerr} for the relative errors. \label{fig:fit1500}}
\end{figure}

\begin{figure}
\centering
\includegraphics[width=0.99\linewidth]{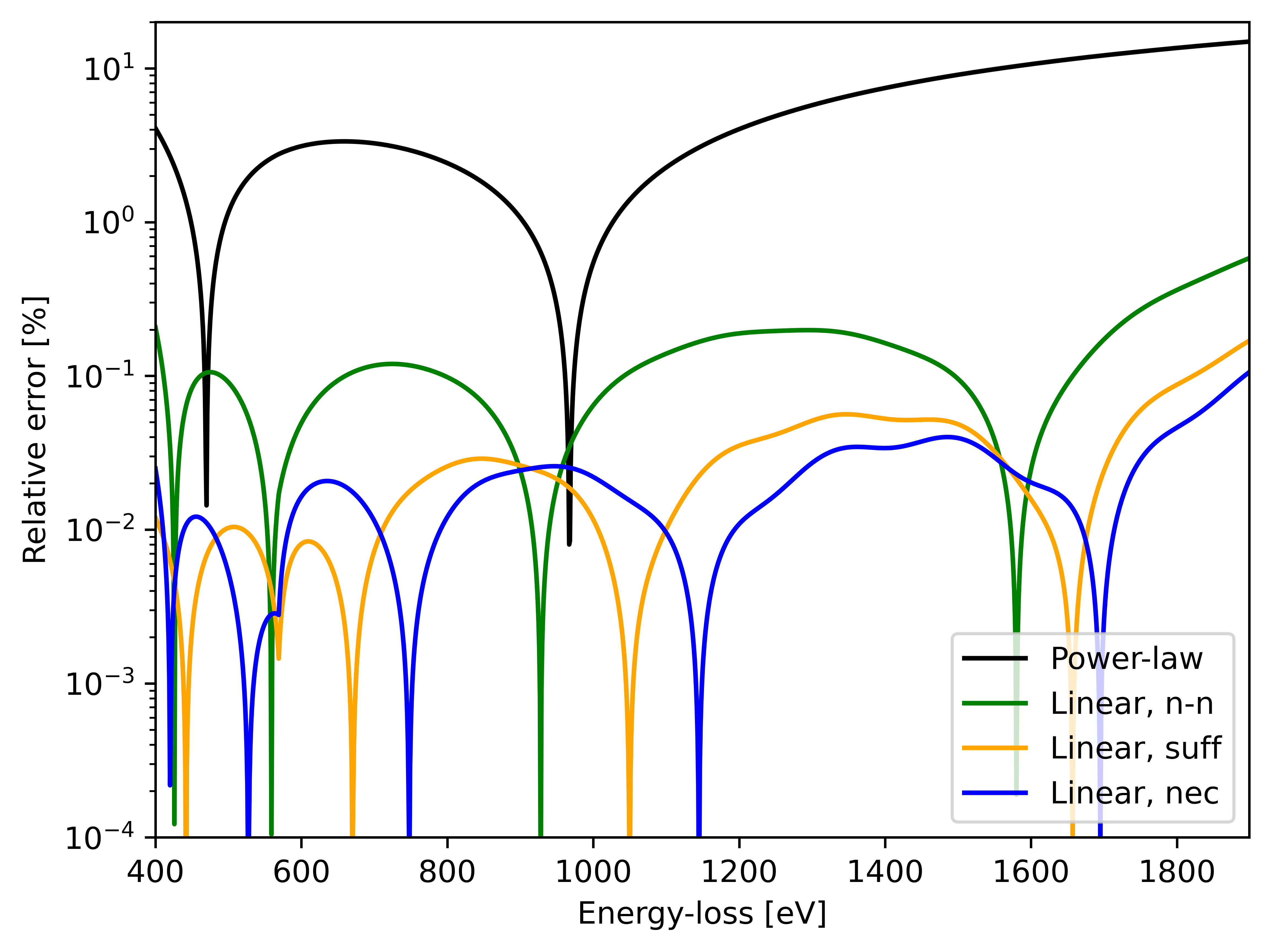}
\caption{Relative errors of fits to the simulated background as a function of energy-loss for the power-law and a five-term linear model with non-negative coefficients (n-n), sufficient (suff), and necessary (nec) constraints. \label{fig:relerr}}
\end{figure}

The results are presented in Table \ref{tab:simbg} and Figures \ref{fig:fit1500} and \ref{fig:relerr}. Over all background lengths, the linear models for all constraints perform notably better than the power-law, and for energy ranges of 500~eV and above, the relative errors of the sufficient and necessary constraints were equal or comparable. For the 1500~eV-wide background, the maximum relative error over this energy range is 15\% for the power-law, while for the linear backgrounds it is 0.6\%, 0.2\% and 0.1\% for the non-negative coefficients, sufficient and necessary constraints, respectively.

These errors can be put in perspective when assuming  counting statistics is the only source of noise. In that case, the deviation from an incorrect model becomes dominant at a total number of counts in the spectrum equal to,
\begin{eqnarray}
  1 / \langle \sigma_{\text{rel}} \rangle^2.    
\end{eqnarray}
Per Table \ref{tab:simbg}, for $\Delta E = 1500$~eV, this amounts to only 242 electrons for the power-law, but 5.0\E+7 for the linear background with sufficient conditions. This demonstrates that for most practical uses, the power law performs largely insufficiently and will lead to significant deviations and bias from the statistical optimum. 

It can be argued that this conclusion depends on the trust we put in how well our simulated background describes reality. If we took the position that, contrary to experimental evidence \cite{egerton1996bg}, a single power-law is considered as the true model for the background, we can still test how the linear background model with sufficient constraints performs when fitted to this power-law over a 1500~eV wide energy range. When the exponent of the power-law is chosen as $-3.5$, we find that a relative error of \mbox{1.8\E-2\%} 
 is obtained showing that, if the power-law were the true model (\emph{quod non}), it would take at least 3.1\E+7 electrons in an experimental spectrum to statistically prefer it over the more generic linear model.

\subsubsection{Experimental backgrounds}

In order to compare the fit quality of the power-law and the linear background model to the experimental backgrounds, the normalized chi-squared, corrected for degrees of freedom \cite{barlow1991},
\begin{eqnarray}
    \langle \chi^2 \rangle = \frac{1}{\text{dof}} \sum_{i=1}^N  \frac{\left(f_i - g_i \right)^2}{f_i}, \label{eq:chisq}
\end{eqnarray}
is used. The variables are defined like in (\ref{eq:relerr}), and $\text{dof}$ is the number of degrees of freedom which equals the number of energy bins minus the number of fitted parameters, thus ensuring that the expected value is 1.

Only sufficient constraints have been investigated, as on the simulated backgrounds they yielded relative errors similar to the necessary constraints, but with the added advantage of guaranteeing convexity. Besides the five-term background with exponents listed in (\ref{eq:ri5}), also four- and three-term backgrounds with exponents
\begin{eqnarray}
  r_1, \ldots, r_4 & = & 1, \ 2.33, \ 3.67, \ 5,  \label{eq:ri4}\\
  r_1, \ r_2, \ r_3 & = & 1, \ 3, \ 5, \label{eq:ri3}
\end{eqnarray}
respectively, were investigated. 

\begin{table}
\center 
\begin{tabular}{ | c | c | c | c | c | c | }
   \hline
   $\Delta E$ & pwr & suff/3 & suff/4 & suff/5 \\
   \hline
150 & 0.99 & 0.98 & 0.98 & 0.98 \\
200 & 1.05 & 1.02 & 1.02 & 1.02 \\
300 & 1.02 & 1.01 & 1.01 & 1.01 \\
500 & 1.06 & 1.00 & 1.00 & 1.00 \\
700 & 1.15 & 1.01 & 1.00 & 1.00 \\
1000 & 1.39 & 1.02 & 1.00 & 1.00 \\
1500 & 2.60 & 1.02 & 1.01 & 1.01 \\
   \hline
\end{tabular}
\caption{Average $\langle \chi^2 \rangle$ of power-law (pwr), and linear model with sufficient conditions and three (suff/3), four (suff/4) and five (suff/5) terms fitted to experimental backgrounds.  See  Table \ref{tab:r} for the number of spectra per energy range.}
\label{tab:expbg} 
\end{table}

\begin{figure}
\centering
\includegraphics[width=0.99\linewidth]{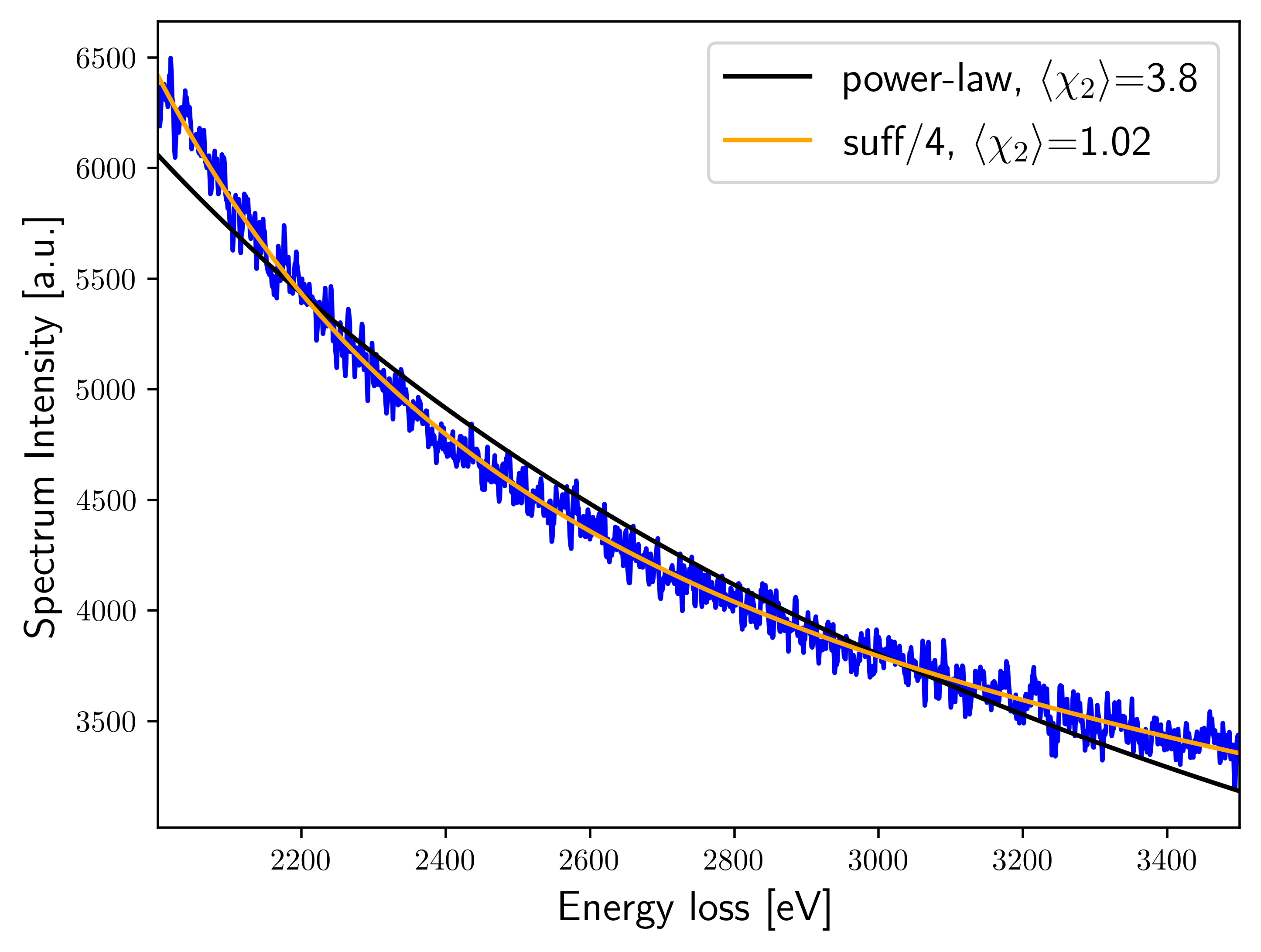}
\caption{Fit of the power-law, and the four-term linear background model with sufficient constraints. \label{fig:fit_examples}}
\end{figure}

The results are listed in Table \ref{tab:expbg}. The linear background models are a significant improvement over the power-law model, especially for larger energy regions. In Figure \ref{fig:fit_examples} an example is shown for a 1500~eV energy range. The poor performance of the power-law model, especially over large energy ranges, is expected from previous work \cite{egerton_1979, Su_1992}.

\section{Experimental results}
\label{sec:expres}

In order to test the performance of the background model in a realistic spectrum imaging situation we selected a semiconductor sample  from a recent advanced device containing N, Ti, O, and Cu. We acquired spectral data with an EEL spectrometer, attached to a Thermo Fisher Themis Z. The acceleration voltage was 300~kV, the convergence and collection semi-angles were 28~mrad and 40~mrad, respectively. The energy axis ranged from 280.384~eV to 1119.193~eV, with a 0.423~eV dispersion. The spectrum image (SI) sampling is 128~by~128 pixels, with a 4.39~nm probe step size.

The edges that were investigated are the C K-edge, N K-edge, Ti L$_{23}$-edge, O K-edge, and Cu L$_{23}$-edge, with nominal onsets at 284~eV, 402~eV, 456~eV, 532~eV, and 931~eV, respectively. 

The spectra in the SI are fitted over the whole energy range with the exception of the interval $[460, 475]$~eV, which was excluded because of the prominent Ti white lines that reside there. The edge models are the hydrogenic K- and L-edges as described in \cite{egerton1996}. Two background models are compared: the power-law model, and the five-term linear model from (\ref{eq:ri5}). 
The low-loss spectra have been recorded as well and are convolved over the models to account for multiple scattering.

A five-term linear background model has been fitted with QP to impose sufficient constraints for convexity, monotonic decrease and non-negativity on the background, and non-negativity on the amplitudes of the edges. Furthermore, the power-law model has been fitted with TRF \cite{trf}, a trust-region reflective method in SciPy that was used to impose non-negativity on the edge amplitudes.

\begin{figure*}
\centering
\includegraphics[width=0.99\linewidth]{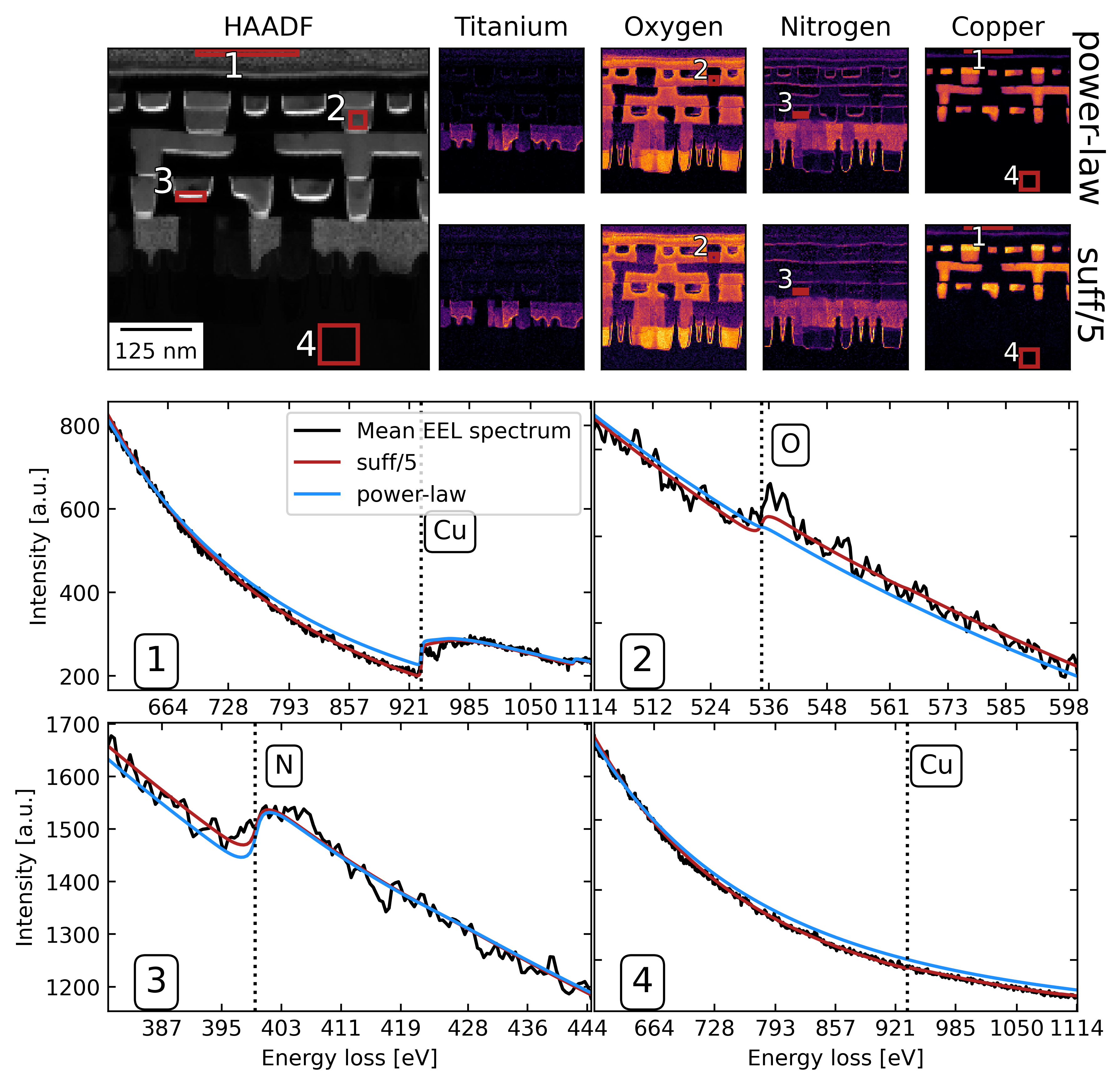}
\caption{Overview of a cross section of a semiconductor sample (HAADF), with four derived elemental maps -- Ti, O, N and Cu -- estimated with a nonlinear power-law model (upper row) and the proposed constrained linear (second row) background (suff/5). The average spectra over four spatial regions (lower two rows) illustrates significant bias for the power law while this is not the case for the constrained linear fit.\label{fig:allmaps}}
\end{figure*}

In Figure \ref{fig:allmaps} elemental maps for N, Ti, O and Cu are displayed. Since each fit is performed on an individual spectrum that is relatively noisy (average 80 counts/energy bin), the quality of the individual fits is not immediately apparent. Hence, we present averages over selected spatial regions, indicated by red squares in the elemental maps, to better assess the results and to show the effect of bias when summing multiple fits.  

In Region 1, in the upper part of the maps, the power-law overestimates the background, resulting in an underestimation of the copper content. The linear background model remains much closer to the average experimental spectrum, except in the fine structure region where we know the atomic cross section model is insufficient. In Region 2, the oxygen edge is not detected with the power-law background, thus missing the slight oxidation that the linear background reveals. The nitrogen content in Region 3 appears overestimated for the power-law model specifically by underestimating the pre-edge region showing bias introduced by an incorrect model. Finally, a systematical overestimation (bias) of the background by the power-law model in the silicon substrate region is demonstrated in Region 4.
As we don't know the true quantification for an unknown experimental sample, it is hard to quantify this bias effect, the example however clearly shows a far lower bias when using the constrained model as compared to the commonly used power law.

\begin{figure*}
\centering
\includegraphics[width=0.99\linewidth]{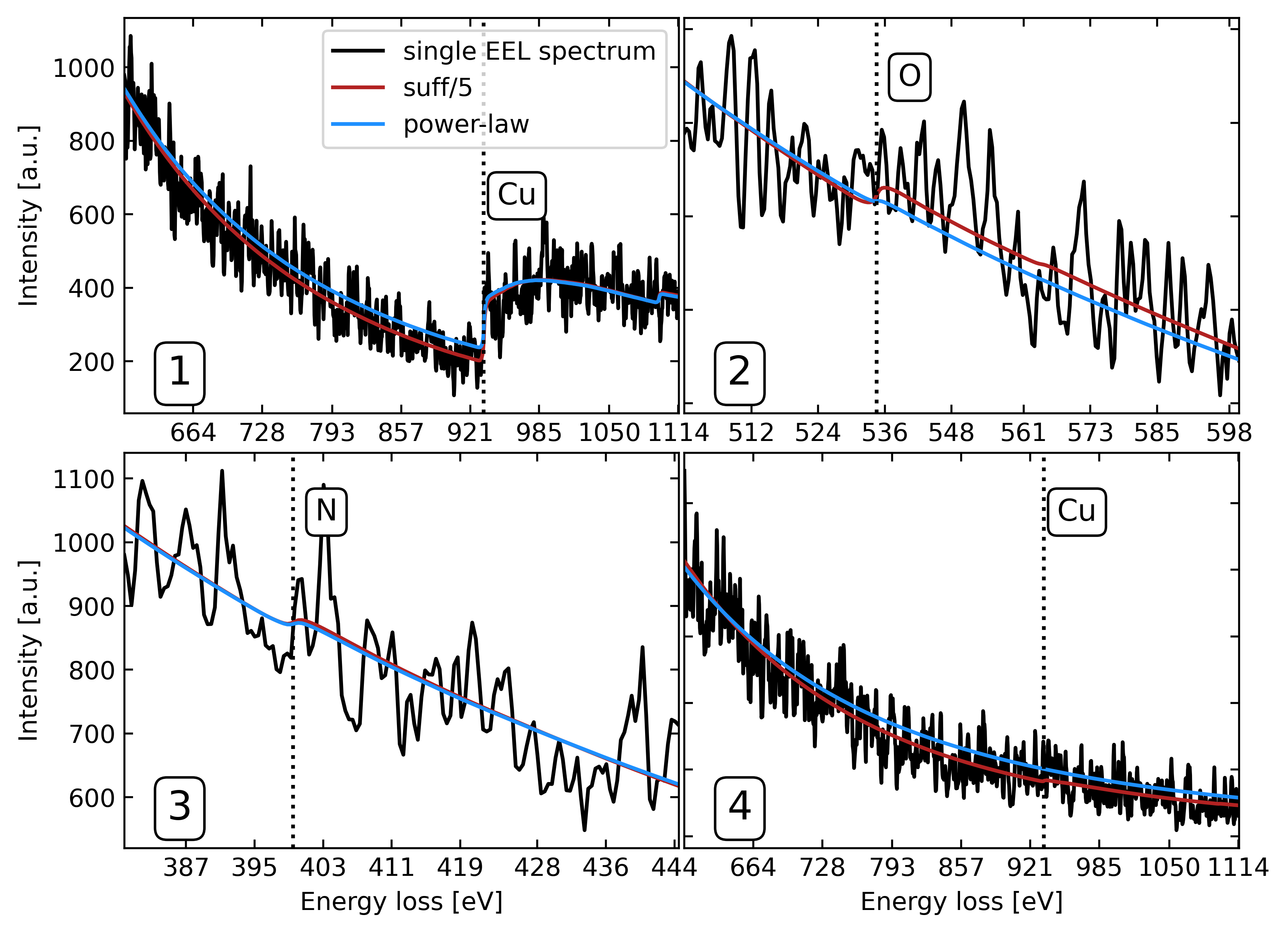}
\caption{Single spectra and their fits, from the centers of the regions depicted in Figure \ref{fig:allmaps}. \label{fig:noisy}}
\end{figure*}

To provide an impression of the noise content of the individual spectra, and the challenges this poses on the fitting algorithms, in Figure \ref{fig:noisy}, a single spectra from the center of each of the four regions in Figure \ref{fig:allmaps} is displayed.

The computational speed of QP (linear constrained model) and TRF (nonlinear power law fit) in our current implementation resulted in 3.6~ms/spectrum for the QP vs 66~ms/spectrum for the TRF. This adds another significant advantage to the linear model being 18 times faster.


\section{Discussion}
\label{sec:dis}

A sum of power-laws is an attractive choice of model, since it lends itself to efficiently incorporate positivity, monotonic decrease and convexity into the solution. As a result, this particular basis (\emph{i.e.} powers of $1/E$) only needs $2^{n-1}+2$ boundary conditions to enforce the constraints. 

Non-convex basis functions, such as polynomials of $E$~\cite{shuman_1985}, might in principle be suitable as well. However it is not clear how to ensure appropriate solutions without applying constraints on each energy bin, which would deteriorate fitting speed and performance. In fact, it is not clear if an attainable set of constraints could be formulated for such basis functions.

The downward spikes in the relative errors depicted in Figure \ref{fig:relerr} coincide with crossings between simulated background and fitted model. Their number loosely correlates with the number of free parameters in the model, since a higher number allows for a better fit. Note that it is not a one-on-one relationship: the fit with non-negative coefficients constraints, for instance, exhibits one fewer zero crossing.

Since solutions with sufficient constraints are a subset of those with necessary constraints, the latter are expected to yield a better fit, as is indeed borne out in Table \ref{tab:expbg}.  However, this comes at the cost of not guaranteeing convexity. Furthermore, this does not preclude the possibility of the relative error for sufficient constraint being lower for specific energies, which can be observed in Figure \ref{fig:relerr}.

Although the spectra in Section \ref{sec:expres} were fitted over the whole energy range, one could argue that by fitting the Cu-edge separately from the other elements over a shorter range, the power-law would have performed better than it did now. We maintain that the capability to treat large energy windows is important nonetheless; for instance, consider that all of the following edges, each of them common in the semiconductor industry, fall in the gap between the O K- and Cu L$_{23}$-edge: Mn-L$_{23}$ at 640~eV, Fe-L$_{23}$ at 708~eV, Co-L$_{23}$ at 779~eV, La-M$_{45}$ at 832~eV, and Ni-L$_{23}$ at 855~eV. The presence of any of these would preclude splitting off the Cu-region.

\section{Conclusion}
\label{sec:con}

A linear background model was proposed for EELS experiments. To prevent artifacts where the fitted background is non-convex, locally increasing or negative, constraints have been formulated that were imposed with the aid of quadratic programming.  The number of constraints equals $2^{n-1} + 2$ at maximum, with $n$ the number of linear parameters. 

The linear model described EELS backgrounds better than the conventional power-law model; this was corroborated by fitting to data sets of simulated and experimental backgrounds and comparing the relative error and the chi-squared to that of the power-law model. Furthermore, energy windows of up to 1500~eV wide were easy to treat as a whole, thus alleviating the need to split up the spectra, something that is not always possible when many different edges are present. This is especially important for large datasets where reducing the amount of user input is attractive and opens opportunities for a fast and entirely unsupervised quantification. We demonstrated this on a semiconductor example dataset which showed a strongly reduced bias resulting in a more reliable quantification.
This is of high importance to alleviate operator bias and reproducibility issues in EELS while at the same time making EELS a far more user friendly technique.


\appendix

\renewcommand{\theequation}{\thesection.\arabic{equation}}
\numberwithin{equation}{section}

\section{Lemma sufficient convexity constraint}
\label{sec:conpro}

\begin{figure}
\centering
\includegraphics[width=0.85\linewidth]{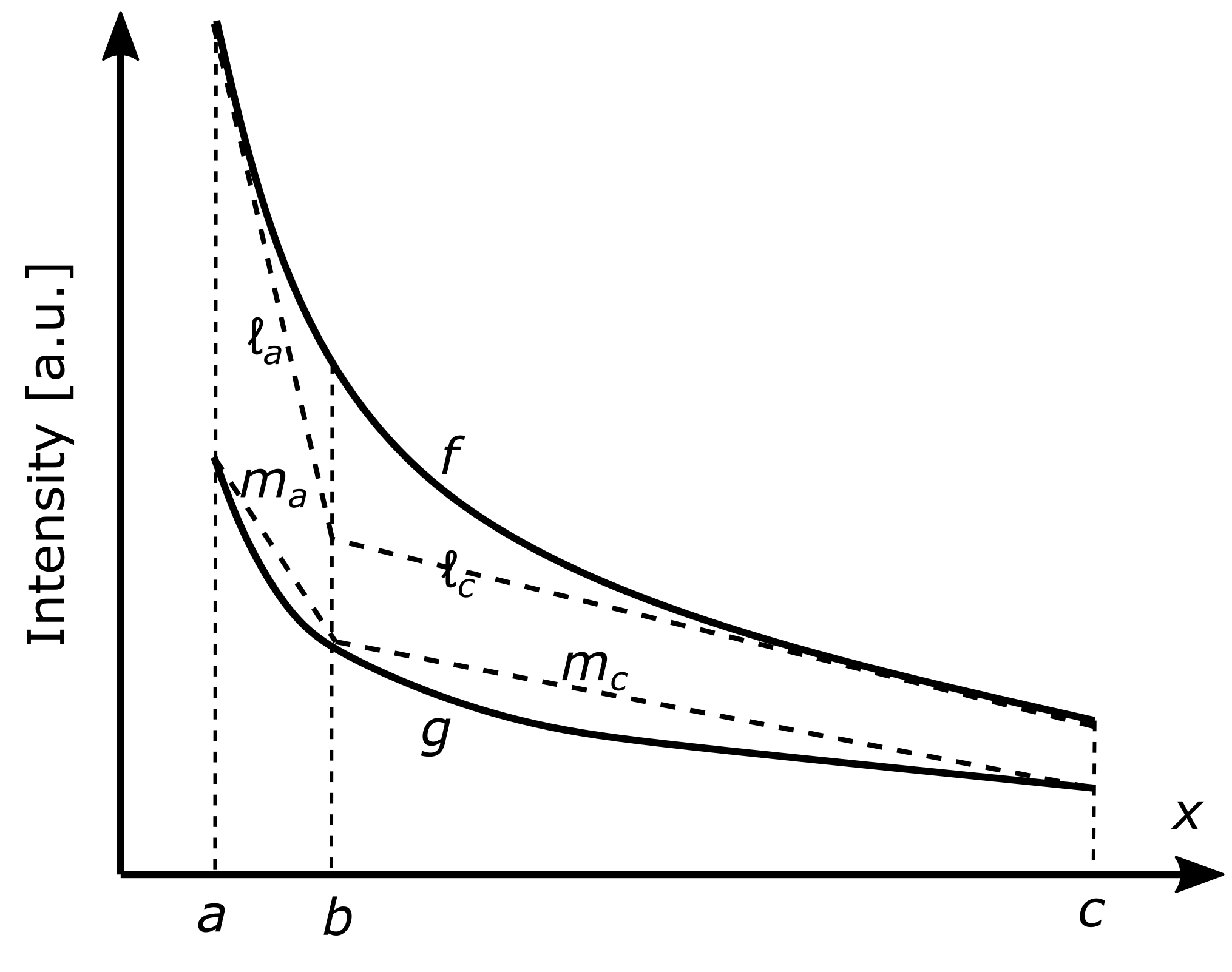}
\caption{The convex function $g(x)$ lies below the convex function $f(x)$ if $g(a) \leq f(a)$, $g(c) \leq f(c)$ and $g(b) \leq \ell_c(b)$. \label{fig:procon}}
\end{figure}

Consider the convex functions $f(x)$ and $g(x)$ in Figure \ref{fig:procon}. We set out to proof that for $g(x)$ to be lower than or equal to $f(x)$ for all $x$ in $[a, c]$, it is sufficient that $g(a) \leq f(a)$, $g(c) \leq f(c)$, and $g(b) \leq \ell_c(b)$; where $\ell_a$ and $\ell_c$ are the tangents of $f$ through $a$ and $c$, respectively, and $b$ is the abscissa of the crossing of $\ell_a$ and $\ell_c$.

Due to $f$'s convexity, the two line segments through the point pair $(a, \ell_a(a))$ and $(b,\ell_a(b))$, and the point pair $(b, \ell_c(b))$ and $(c, \ell_c(c))$ are lower than or equal to $f(x)$ for all $x$ in the open interval $(a, c)$. It is hence sufficient to prove that $g(x)$ is lower than or equal to the two line segments if the above conditions are fulfilled.

Consider the line segment $m_a$ through the points $(a, g(a))$ and $(b,g(b))$. Due to $g$'s convexity, $g(x) \leq m_a(x)$ for all $x$ in $[a,b]$, and since it holds that $m_a(a) \leq \ell_a(a)$ and $m_a(b) \leq \ell_a(b)$, it follows from the line segments' linearity that $ m_a(x) \leq \ell_a(x)$, and hence $g(x) \leq \ell_a(x)$, for all $x$ in $[a,b]$. The same reasoning applies, \emph{mutatis mutandis}, for the interval $[b,c]$. 

Note how the proof is independent of explicit expressions for $f$ and $g$ and holds for all positive, decreasing and convex basis functions, e.g. $\exp \left( -E/s \right)$, with $s$ a positive constant.

\section{Looser sufficient convexity constraints}
\label{sec:loose}

\begin{figure}
\centering
\includegraphics[width=0.85\linewidth]{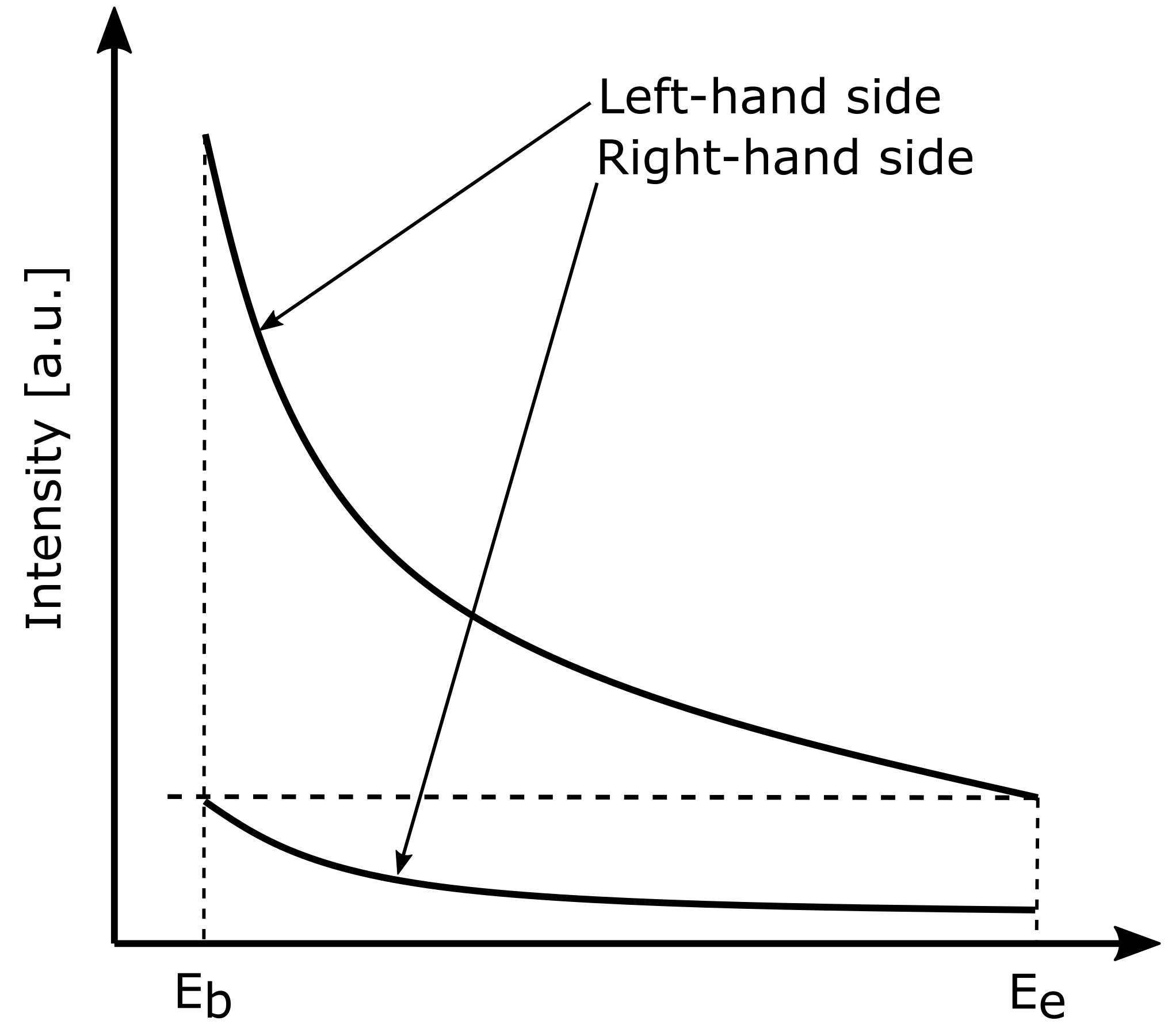}
\caption{Illustration of the convexity constraints in (\ref{eq:loosest}). \label{fig:loosest}}
\end{figure}

\begin{figure}
\centering
\includegraphics[width=0.85\linewidth]{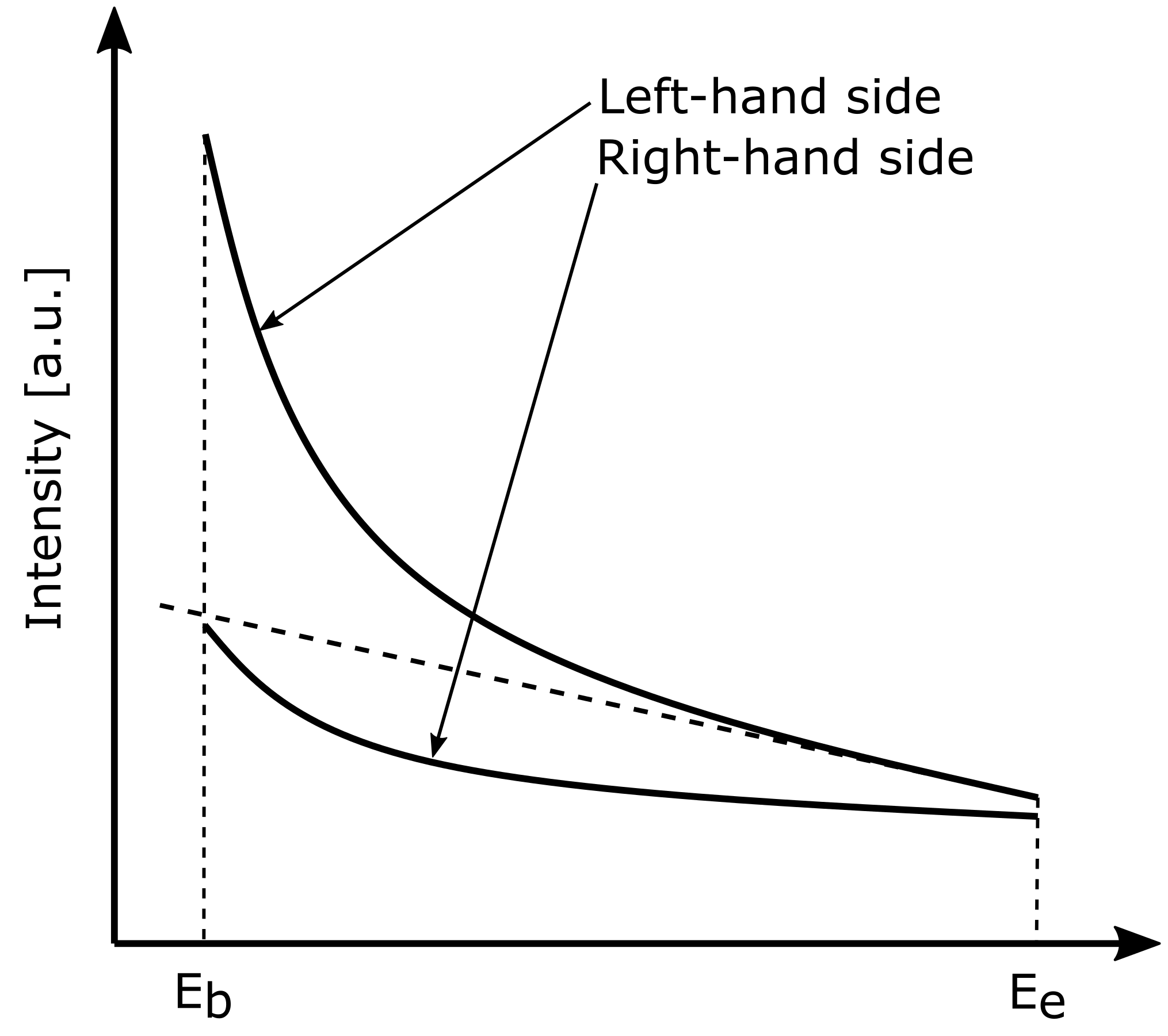}
\caption{Illustration of the convexity constraints in (\ref{eq:lessloose}). \label{fig:lessloose}}
\end{figure}

In this Section sufficient conditions for convexity are shown that are looser than those in the main body of the paper, this means although they guarantee convexity, they exclude more valid solutions. Our starting point is once again (\ref{eq:split}).

\subsection{Loosest conditions}

Equation (\ref{eq:split}) holds for all energies if the right-hand side evaluated in $E_b$ is below the left-hand side evaluated in $E_e$. This is illustrated in Figure \ref{fig:loosest}.  This leads to the set of constraints,
\begin{eqnarray}
  \begin{split}
  & a_1 s_1 + \sum_{i \in \mathcal{C}_k} a_i s_i E_b^{\Delta r_i} \\
  & + \sum_{i \notin \mathcal{C}_k} a_i s_i E_e^{\Delta r_i} \geq 0, \ \forall \mathcal{C}_k,
  \end{split} \label{eq:loosest}
\end{eqnarray}
with $\mathcal{C}_k$ defined like in Section \ref{sec:sufcon}.

\subsection{Less loose conditions}

Equation (\ref{eq:split}) holds for all energies if (\emph{i}) the right-hand side is below the left-hand side in $E_b$, and  (\emph{ii}) the right-hand side evaluated in $E_b$ is below the left-hand side's tangent through $E_e$ evaluated in $E_b$. This is illustrated in Figure \ref{fig:lessloose}.  This leads to the set of constraints,
\begin{eqnarray}
  \sum_i a_i s_i E_e^{\Delta r_i} \geq 0, \quad & \\
  \begin{split}
  &\sum_{i \in \mathcal{C}_k} a_i s_i E_e^{\Delta r_i} \left( 1 + (E_b/E_e - 1) \Delta r_i \right) \\
  & + a_1 s_1 + \sum_{i \notin \mathcal{C}_k} a_i s_i E_b^{\Delta r_i} \geq 0, \ \forall \mathcal{C}_k,
  \end{split} \label{eq:lessloose}
\end{eqnarray}
with $\mathcal{C}_k$ defined like before.

\section*{Acknowledgements}

This project has received funding from the ECSEL Joint Undertaking (JU) under grant agreement No 875999. The JU receives support from the European Union’s Horizon 2020 research and innovation programme and Netherlands, Belgium, Germany, France, Austria, Hungary, United Kingdom, Romania, Israel

The data is available upon request.

\FloatBarrier

\small

\bibliographystyle{unsrt}
\bibliography{references}

\end{document}